\documentclass[showpacs,aps,prl,twocolumn,superscriptaddress,psfrag,amsmath,amssymb,longbibliography]{revtex4-1}
\usepackage[pdftex]{graphicx} 
\usepackage{epstopdf}
\usepackage{verbatim}
\usepackage{amsmath}
\usepackage{color}
\usepackage{subfigure}
\usepackage{amsbsy}
\usepackage{wasysym}
\usepackage{textcomp}
\usepackage{times}
\usepackage{float}
\usepackage{latexsym,amsmath,amssymb,bm,euscript}
\usepackage[colorlinks=true,linkcolor=blue,citecolor=blue]{hyperref}
\usepackage{hyperref}
\usepackage{soul}
\usepackage[normalem]{ulem}
\usepackage{mathrsfs}
\usepackage{lettrine}
\usepackage{xspace}
\usepackage{filecontents}

\begin{document}

\title{Spiral spin liquid in a frustrated honeycomb antiferromagnet: A single-crystal study of GdZnPO}

\author{Zongtang Wan}
\thanks{These authors contributed equally to this work}
\affiliation{Wuhan National High Magnetic Field Center and School of Physics, Huazhong University of Science and Technology, 430074 Wuhan, China}
\author{Yuqian Zhao}
\thanks{These authors contributed equally to this work}
\affiliation{Wuhan National High Magnetic Field Center and School of Physics, Huazhong University of Science and Technology, 430074 Wuhan, China}
\author{Xun Chen}
\affiliation{Wuhan National High Magnetic Field Center and School of Physics, Huazhong University of Science and Technology, 430074 Wuhan, China}
\author{Zhaohua Ma}
\affiliation{Wuhan National High Magnetic Field Center and School of Physics, Huazhong University of Science and Technology, 430074 Wuhan, China}
\author{Zikang Li}
\affiliation{Wuhan National High Magnetic Field Center and School of Physics, Huazhong University of Science and Technology, 430074 Wuhan, China}
\author{Zhongwen Ouyang}
\affiliation{Wuhan National High Magnetic Field Center and School of Physics, Huazhong University of Science and Technology, 430074 Wuhan, China}
\author{Yuesheng Li}
\email{yuesheng\_li@hust.edu.cn}
\affiliation{Wuhan National High Magnetic Field Center and School of Physics, Huazhong University of Science and Technology, 430074 Wuhan, China}

\begin{abstract}
The frustrated honeycomb spin model can stabilize a subextensively degenerate spiral spin liquid with nontrivial topological excitations and defects, but its material realization remains rare. Here, we report the experimental realization of this model in the structurally disorder-free compound GdZnPO. Using a single-crystal sample, we find that spin-7/2 rare-earth Gd$^{3+}$ ions form a honeycomb lattice with dominant second-nearest-neighbor antiferromagnetic and first-nearest-neighbor ferromagnetic couplings, along with easy-plane single-site anisotropy. This frustrated model stabilizes a unique spiral spin liquid with a degenerate contour around the $\mathrm{K}$$\{$1/3,1/3$\}$ point in reciprocal space, consistent with our experiments down to 30 mK, including the observation of a giant residual specific heat. Our results establish GdZnPO as an ideal platform for exploring the stability of spiral spin liquids and their novel properties, such as the emergence of low-energy topological defects on the sublattices.
\end{abstract}

\maketitle

\emph{Introduction.}---In magnets, frustration can lead to the emergence of exotic phases such as spin liquids, which exhibit fractionalized excitations and topological order~\cite{balents2010spin,savary2016quantum,zhou2017quantum}. These phases intimately relate to significant scientific challenges, including understanding high-temperature superconductivity~\cite{anderson1987resonating} and realizing topological quantum computation~\cite{nayak2008non}. Frustrated magnetic interactions on lattices like triangular, kagome, and pyrochlore can result in extensively degenerate ground states with finite zero-point entropy per spin proportional to $\ln(2S+1)$~\cite{wannier1950antiferromagnetism,kano1953antiferromagnetism,ramirez1999zero}, where $S$ is the spin quantum number. Such spin liquids can transition between different ground states by flipping only a few spins, reflecting a local nature of ground-state degeneracy~\cite{PhysRevResearch.4.023175}. In contrast, a spiral spin liquid (SSL) fluctuates cooperatively among subextensively degenerate spiral configurations, with ground-state wave vectors ($\mathbf{Q}_\mathrm{G}$) forming a continuous contour or surface in reciprocal space for two- (2D) or three-dimensional (3D) systems, respectively~\cite{yao2021generic}. Transitioning between these configurations requires a global action on all spins~\cite{PhysRevResearch.4.023175}. Moreover, SSLs possess much lower ground-state degeneracy, with zero-point entropy per site proportional to $\ln(2S+1)/N^{1/d_0}\rightarrow0$ in the thermodynamic limit, where $N$ is the number of spins and $d_0$ is the system's dimension.

SSLs give rise to topological excitations/defects, such as spin vortices~\cite{PhysRevB.93.085132,gao2017spiral} and momentum vortices~\cite{PhysRevResearch.4.023175,PhysRevB.110.085106}, which can be utilized in antiferromagnetic spintronics without leakage of magnetic fields~\cite{jungwirth2016antiferromagnetic,RevModPhys.90.015005,gao2020fractional}, topologically protected quantum memory and logic operations~\cite{yao2013topologically}, and the study of fracton gauge theory~\cite{PhysRevB.95.115139,nandkishore2019fractons,pretko2020fracton}, among other applications. These defects also play a crucial role in determining the low-energy properties of the system~\cite{PhysRevResearch.4.023175}. SSLs on a frustrated honeycomb lattice may induce topological magnon bands and unconventional thermal Hall effects~\cite{owerre2017topological,PhysRevB.106.035113}. Moreover, SSLs are proposed as a possible route to achieving the long-sought quantum spin liquids by tuning $S$ from the classical ($\rightarrow$ $\infty$) to the quantum (= 1/2) limit~\cite{niggemann2019classical}. Despite extensive theoretical efforts, the explicit realization of SSLs in real materials remains rare, with only a few examples reported, such as the 3D diamond-lattice MnSc$_2$S$_4$~\cite{bergman2007order,gao2017spiral,PhysRevB.98.064427,gao2020fractional} and LiYbO$_2$~\cite{PhysRevLett.130.166703}, the 3D buckled honeycomb Cs$_3$Fe$_2$Cl$_9$~\cite{gao2024,PhysRevB.103.104433}, the quasi-2D bilayer breathing-kagome Ca$_{10}$Cr$_7$O$_{28}$~\cite{PhysRevB.104.024426,takahashi2024}, and the quasi-2D honeycomb FeCl$_3$~\cite{PhysRevLett.128.227201}. Moreover, most existing theoretical and experimental studies have focused on SSL systems with $\mathbf{Q}_\mathrm{G}$ around the $\Gamma$$\{$0,0$\}$ point, leaving the (topological) properties of SSLs with $\mathbf{Q}_\mathrm{G}$ around non-$\Gamma$ points still unclear.

SSLs in the 2D $J_1$-$J_2$ frustrated honeycomb system can exist over a wide range of interaction parameters, 1/2 $>$ $|J_2/J_1|$ $>$ 1/6 ($\mathbf{Q}_\mathrm{G}$ around the $\Gamma$ point) and $|J_2/J_1|$ $>$ 1/2 ($\mathbf{Q}_\mathrm{G}$ around the $\mathrm{K}$$\{$1/3,1/3$\}$ point)~\cite{PhysRevB.81.214419,owerre2017topological,PhysRevB.106.035113,okumura2010novel,PhysRevB.100.224404,PhysRevResearch.4.013121}. However, their experimental realization has only been explicitly reported in FeCl$_3$ to date~\cite{PhysRevLett.128.227201}. In FeCl$_3$, the honeycomb lattice is formed by transition-metal Fe$^{3+}$ ($S$ = 5/2) ions, with an interlayer separation of $c$/3 $\sim$ 5.8 \AA, comparable to the second-nearest-neighbor intralayer distance $a$ $\sim$ 6.1 \AA. Due to the spatial delocalization of the 3$d$ electrons, the interlayer couplings are non-negligible compared to $J_2$, complicating many-body modeling~\cite{PhysRevLett.128.227201}. The spiral contour observed around the $\Gamma$ point in FeCl$_3$ aligns with the ratio $|J_2/J_1|$ $\sim$ 0.36~\cite{PhysRevLett.128.227201}, but the honeycomb SSL with $\mathbf{Q}_\mathrm{G}$ around the $\mathrm{K}$ point has yet to be realized.

\begin{figure}
\begin{center}
  \includegraphics[width=8.7cm,angle=0]{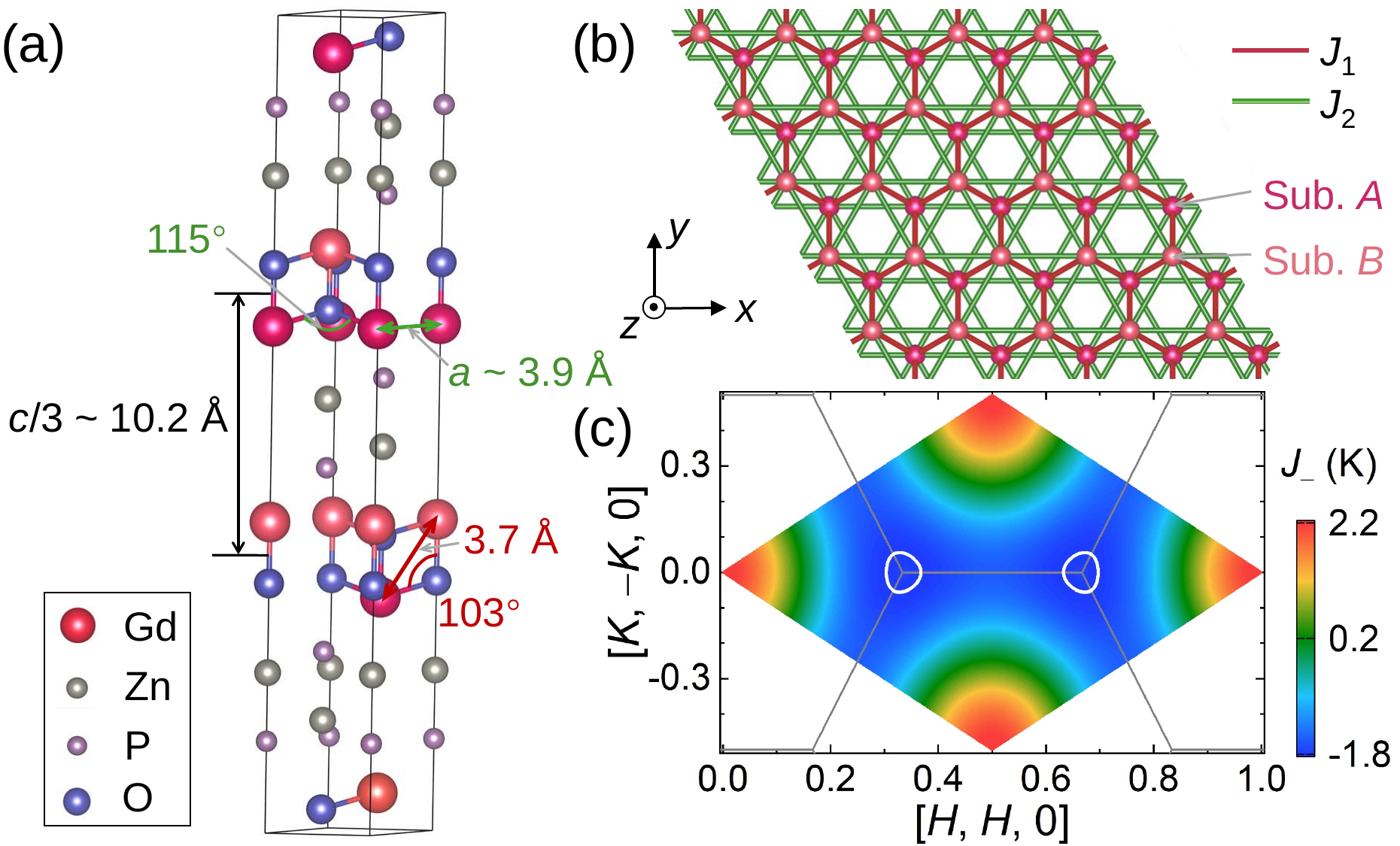}
  \caption{(a) The GdZnPO crystal structure and (b) the honeycomb lattice in the $ab$ plane formed by Gd$^{3+}$ ions. The inset in (b) defines the coordinate system for the spin components. (c) Wave-vector dependence of the lower eigenvalue of the interaction matrix, $J_-$, calculated from the GdZnPO spin Hamiltonian. The white lines indicate the spiral contour where $J_-$ is minimized, and the grey lines mark Brillouin zone boundaries.}
  \label{fig1}
\end{center}
\end{figure}

To approach the ideal $J_1$-$J_2$ honeycomb model, we investigate the rare-earth Gd$^{3+}$ ($S$ = 7/2, orbital quantum number $L$ = 0) based honeycomb-lattice (or bilayer-triangular-lattice) antiferromagnet GdZnPO (Fig.~\ref{fig1}), which has been successfully synthesized~\cite{nientiedt1998equiatomic,lincke2008magnetic} but remains understudied in terms of low-energy magnetism. Using a single-crystal sample, we determine the exchange Hamiltonian of GdZnPO, with first- and second-nearest-neighbor magnetic couplings $J_1$ $\sim$ $-$0.39 K and $J_2$ $\sim$ 0.57 K, respectively, and an easy-plane anisotropy $D$ $\sim$ 0.30 K. Given $|J_2/J_1|$ $\sim$ 1.5, a SSL with $\mathbf{Q}_\mathrm{G}$ around the $\mathrm{K}$ point is expected to form, consistent with the observations of large magnetic specific heat and susceptibility in the low-$T$ limit. Monte Carlo (MC) simulations for GdZnPO reveal the emergence of well-defined spin vortices and antivortices, associated with local (sublattice) momentum vortices, on the sublattices. Future studies may explore Kitaev quantum spin liquids~\cite{kitaev2006anyons} and other quantum phases in structural siblings of GdZnPO with effective $S$ = 1/2 rare-earth ions instead.

\emph{Methods.}---High-quality single crystals of GdZnPO 
were grown using a two-step method with NaCl/KCl flux. Magnetization and specific heat measurements were conducted between 1.8 and 300 K using magnetic and physical property measurement systems (Quantum Design). First-derivative electron spin resonance (ESR) spectra were obtained at 300 K and x-band frequencies ($\sim$ 9.8 GHz) using a continuous-wave spectrometer (Bruker EMXmicro-6/1). Pulsed-field ESR spectra were measured at 97 GHz down to $\sim$ 2 K at Wuhan National High Magnetic Field Center. Specific heat down to 53 mK and magnetization down to 30 mK were measured in a $^3$He-$^4$He dilution refrigerator. Standard MC simulations were conducted by slowly annealing from high to low temperatures on 2$\times L_N^2$ clusters with periodic boundary conditions, where 72 $\geq$ $L_N$ $\geq$ 9~\cite{supple}.

\emph{Exchange Hamiltonian.}---The crystal structure of GdZnPO, determined by x-ray diffraction (XRD)~\cite{supple}, is shown in Fig.~\ref{fig1}(a). We propose GdZnPO as an ideal candidate for the frustrated honeycomb model for the following reasons: (1) Due to the highly localized nature of the 4$f$ electrons~\cite{PhysRevLett.115.167203}, it is sufficient to consider only the nearest-neighbor interactions, $J_1$ (with $|$Gd-Gd$|$ $\sim$ 3.7 \AA) and $J_2$ (with $|$Gd-Gd$|$ $\sim$ 3.9 \AA), while longer-range interactions with $|$Gd-Gd$|$ $\geq$ 5.4 \AA~are negligible ($\lesssim$ 0.016 K). The nearest-neighbor exchanges $J_1$ and $J_2$ are mediated by O$^{2-}$ ions with the bond angles $\angle$Gd-O-Gd = 103$^\circ$ and 115$^\circ$, respectively, indicating $J_2$ $>$ $J_1$. (2) The GdO honeycomb layers are spatially separated by a nonmagnetic ZnP bilayer, with an average distance of $c$/3 $\sim$ 10.2 \AA, suggesting a strong 2D nature of the spin system. (3) Gd$^{3+}$ has the maximum $S$ = 7/2 among all existing SSL candidates, bringing the GdZnPO system closer to the classical prototype. (4) Gd$^{3+}$ is the only magnetic rare-earth ion with $L$ = 0, making spin-spin interaction anisotropy due to spin-orbital coupling negligible. (5) The spatial isotropy of the honeycomb lattice is ensured by the $R\bar{3}m$ space group symmetry of GdZnPO [Fig.~\ref{fig1}(b)]. (6) No site-mixing structural disorder has been reported~\cite{nientiedt1998equiatomic,lincke2008magnetic}, and our refinement of the single-crystal XRD data confirms this. Consequently, the notorious interaction randomness and resulting modeling complexity~\cite{PhysRevLett.118.107202,PhysRevX.10.011007,PhysRevB.105.024418} are not applicable to GdZnPO. (7) The spatial localization of the 4$f$ electrons weakens $J_1$ and $J_2$, allowing the GdZnPO spin system to be fully polarized by an achievable field ($\leq$ 14 T), enabling comprehensive magnetic investigations. (8) Single-crystal samples are available, enabling measurements of spin anisotropy.

\begin{figure}
\begin{center}
  \includegraphics[width=8.75cm,angle=0]{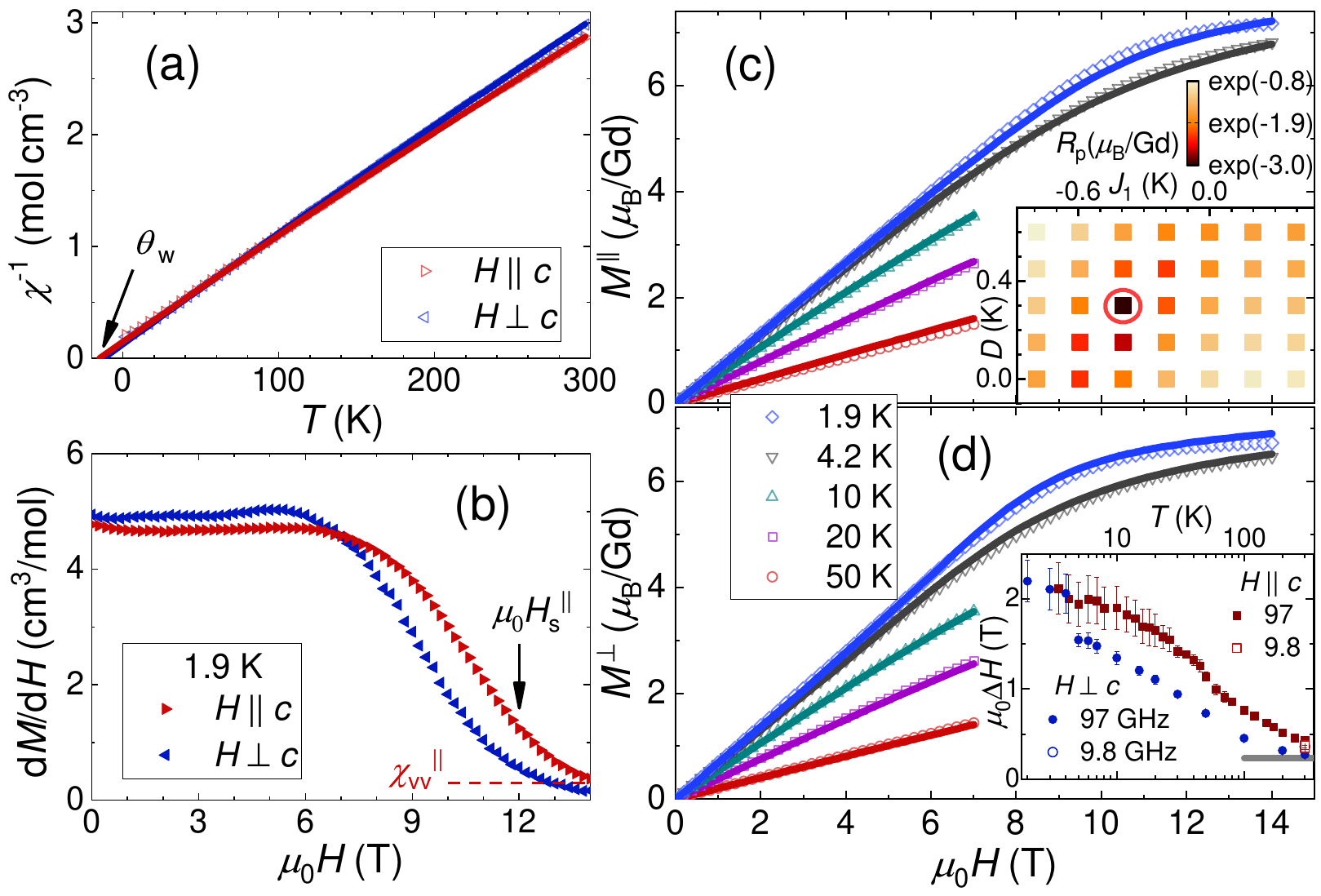}
  \caption{Magnetic properties of GdZnPO. (a) Inverse magnetic susceptibilities measured at 1 T, with colored lines showing Curie-Weiss fits above $\sim$ 20 K. (b) Field dependence of susceptibility (d$M$/d$H$) measured at 1.9 K. Combined MC fit to magnetization at various temperatures in fields parallel (c) and perpendicular (d) to the $c$ axis. Inset of (c): Deviation $R_\mathrm{p}$ of experimental magnetization from theoretical values, with fixed $\theta_\mathrm{w}$ [$\equiv$ $-S(S+1)(J_1+2J_2)$] = $-$12 K. The red circle indicates the optimal parameters, $J_1$ $\sim$ $-$0.4 K and $D$ $\sim$ 0.3 K. Inset of (d): Temperature dependence of ESR linewidths, with the gray line representing the calculated high-temperature limit.}
  \label{fig2}
\end{center}
\end{figure}

The GdZnPO crystals are transparent insulators. 
The generic exchange Hamiltonian is given by
\begin{multline}
\mathcal{H} = J_1\sum_{\langle j0,j1\rangle}\mathbf{S}_{j0}\cdot\mathbf{S}_{j1}+J_2\sum_{\langle\langle j0,j2\rangle\rangle}\mathbf{S}_{j0}\cdot\mathbf{S}_{j2}+D\sum_{j0}(S_{j0}^z)^2\\
-\mu_0H^\parallel g\mu_\mathrm{B}\sum_{j0}S_{j0}^z,
\label{eq1}
\end{multline}
where $D$ represents the single-site anisotropy for ions with $S$ $>$ 1/2~\cite{liu2021frustrated}. At high temperatures ($T$ $\gtrsim$ 20 K), the susceptibilities measured both parallel and perpendicular to the $c$ axis are well fitted by the Curie-Weiss function [Fig.~\ref{fig2}(a)], $\chi$ = $\mu_0N_\mathrm{A}\mu_\mathrm{B}^2g^2S(S+1)/[3k_\mathrm{B}(T-\theta_\mathrm{w})]+\chi_\mathrm{vv}$, where $\chi_\mathrm{vv}$ represents the nearly $T$-independent Van Vleck susceptibility [see Fig.~\ref{fig2}(b)]. We obtained a Weiss temperature $\theta_\mathrm{w}$ = $-$12(1) K and $g$ = 2.01(2). The measured $g$ factor is isotropic and close to the free electron value, confirming the absence of spin-orbital coupling due to $L$ = 0 in GdZnPO. A $g$ $\sim$ 2 value is also detected in our ESR measurements~\cite{supple}.

At low temperatures and low fields, the longitudinal susceptibility $\chi^\parallel$ is smaller than $\chi^\perp$. At higher fields, the spin system is more easily polarized along the $ab$ plane than along the $c$ axis, as shown in Fig.~\ref{fig2}(b). These observations indicate an easy-plane nature of the single-site anisotropy, with $D$ $>$ 0. From a combined fit to the magnetization data by minimizing $R_\mathrm{p}$ = $\sqrt{\sum_j(M_j^\mathrm{exp}-M_j^\mathrm{cal})^2/N_M}$ [Fig.~\ref{fig2}(c) and \ref{fig2}(d)], we obtained $J_1$ $\sim$ $-$0.39(2) K, $J_2$ $\sim$ 0.57(2) K, and $D$ $\sim$ 0.30(2) K for GdZnPO, which forms the core findings of this work. Here, $M_j^\mathrm{exp}$ and $M_j^\mathrm{cal}$ represent the experimental and calculated magnetizations, respectively, and $N_M$ is the number of data points. By fixing $\theta_\mathrm{w}$ $\equiv$ $-S(S+1)(J_1+2J_2)$ to the experimental value of $-$12 K, we mapped $R_\mathrm{p}$, which takes a minimum of 0.05 $\mu_\mathrm{B}$/Gd at the optimized values of $J_1$ and $D$ [see the inset of Fig.~\ref{fig2}(c)]. Using the refined Hamiltonian, we calculated the ESR linewidth along the $c$ axis in the high-$T$ limit, $\mu_0\Delta H$ ($T\rightarrow\infty$) = $\sqrt{64\pi D^4/(21J_1^2+42J_2^2+344D^2/63)}/(g\mu_\mathrm{B})$~\cite{PhysRevB.4.38,PhysRevLett.101.026405}, which yielded $\sim$ 0.23 T, agreeing with the experimental values [see the inset of Fig.~\ref{fig2}(d)].

\emph{Spiral spin-liquid ansatz.}---We calculated the 2$\times$2 interaction matrix via the Fourier transform of Eq.~(\ref{eq1})~\cite{PhysRevB.43.865,paddison2024}.
Diagonalizing this matrix yield two eigenvalues, with the lower one, $J_-$($\mathbf{Q}$), presented in Fig.~\ref{fig1}(c). $J_-$ reaches its minimum along the spiral contour at $\mathbf{Q}_\mathrm{G}$ = $h_\mathrm{G}\mathbf{b}_1$+$k_\mathrm{G}\mathbf{b}_2$, where $\cos(2\pi h_\mathrm{G})+\cos(2\pi k_\mathrm{G})+\cos(2\pi h_\mathrm{G}+2\pi k_\mathrm{G}) = \frac{1}{2}(\frac{J_1^2}{4J_2^2}-3)$~\cite{supple}. Here, $\mathbf{b}_1$ and $\mathbf{b}_2$ are the reciprocal vectors.


The spiral contour remains independent of the longitudinal field below the fully-polarized value, $\mu_0H_\mathrm{s}^\parallel$ = $S[2D+3J_1+9J_2+J_1^2/(4J_2)]/(g\mu_\mathrm{B})$ ($\sim$ 12 T), indicating the stability of the low-$T$ SSL phase against $H^\parallel$ in GdZnPO. Moreover, the ratio $J_2/|J_1|$ $\sim$ 1.5 approaches the $J_2/|J_1|$ $\to$ $\infty$ limit, where $J_2$ promotes a decoupled 120$^\circ$ coplanar order. Consequently, the spiral contour becomes nearly circular with a radius of $|J_1|$/(4$\pi J_2$)~\footnote{Here, the lengths of the reciprocal vectors were set as unit, with $|\mathbf{b}_1|$ = $|\mathbf{b}_2|$ = 1.} around the $\mathrm{K}$ point, suggesting that the effective U(1) symmetry in reciprocal space may still stand. GdZnPO is thus a promising platform for exploring the fracton gauge theory~\cite{PhysRevB.95.115139,nandkishore2019fractons,pretko2020fracton}, potentially after decomposing the honeycomb lattice into six sublattices (as discussed below).
\begin{figure}
\begin{center}
  \includegraphics[width=7.5cm,angle=0]{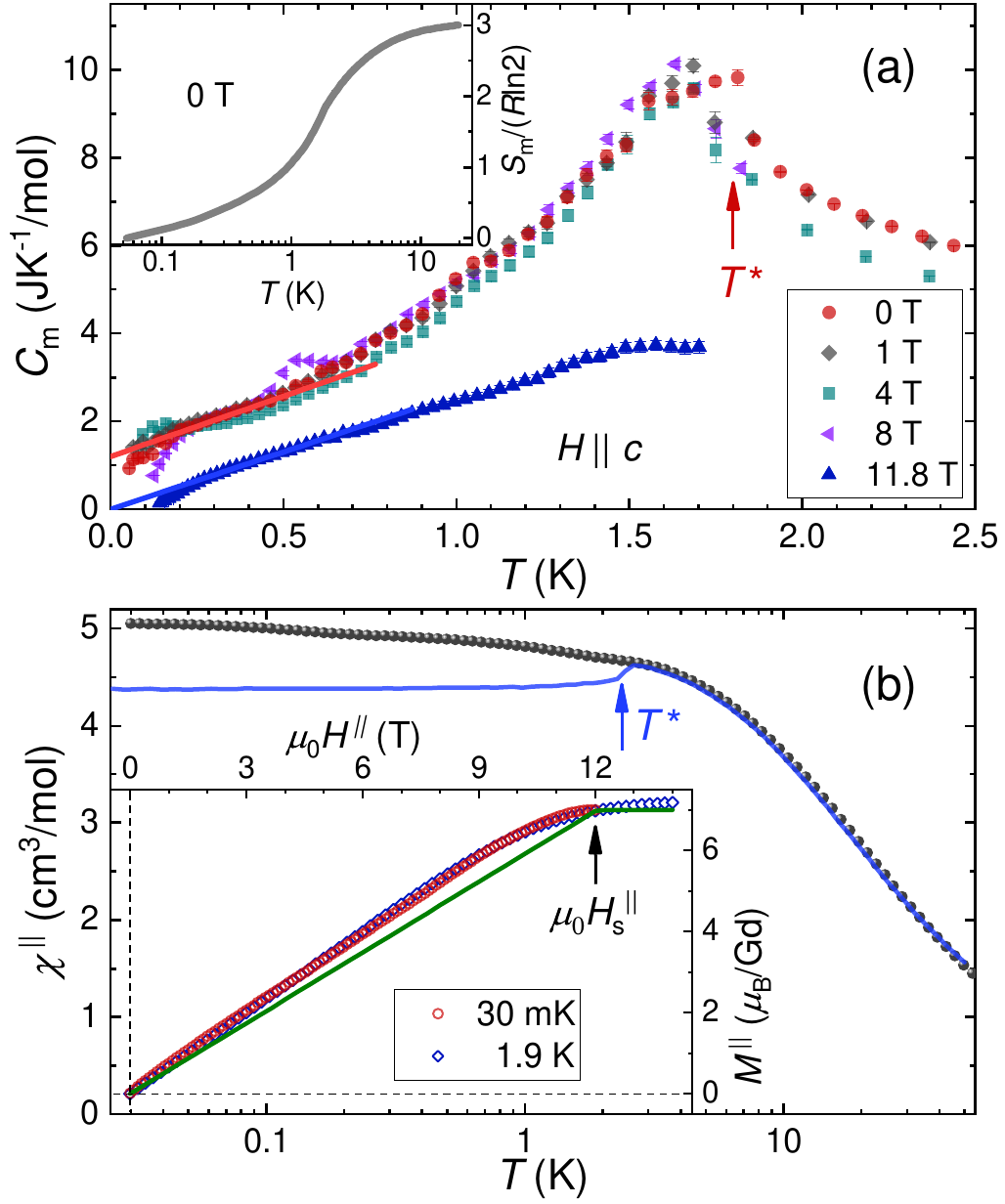}
  \caption{Low-temperature properties of GdZnPO. (a) Temperature dependence of magnetic specific heat under various fields along the $c$ axis. The red and blue lines are guides to the eye; the inset shows magnetic entropy at 0 T. (b) Temperature dependence of magnetic susceptibility ($M^\parallel$/$H^\parallel$) measured with $\mu_0H^\parallel$ = 1 T along the $c$ axis. The line represents the MC simulation. Inset of (b): Longitudinal magnetization measured at 30 mK and 1.9 K, with the green line showing the 0 K calculation using the classical SSL ansatz.}
  \label{fig3}
\end{center}
\end{figure}

\begin{figure*}
\begin{center}
  \includegraphics[width=16cm,angle=0]{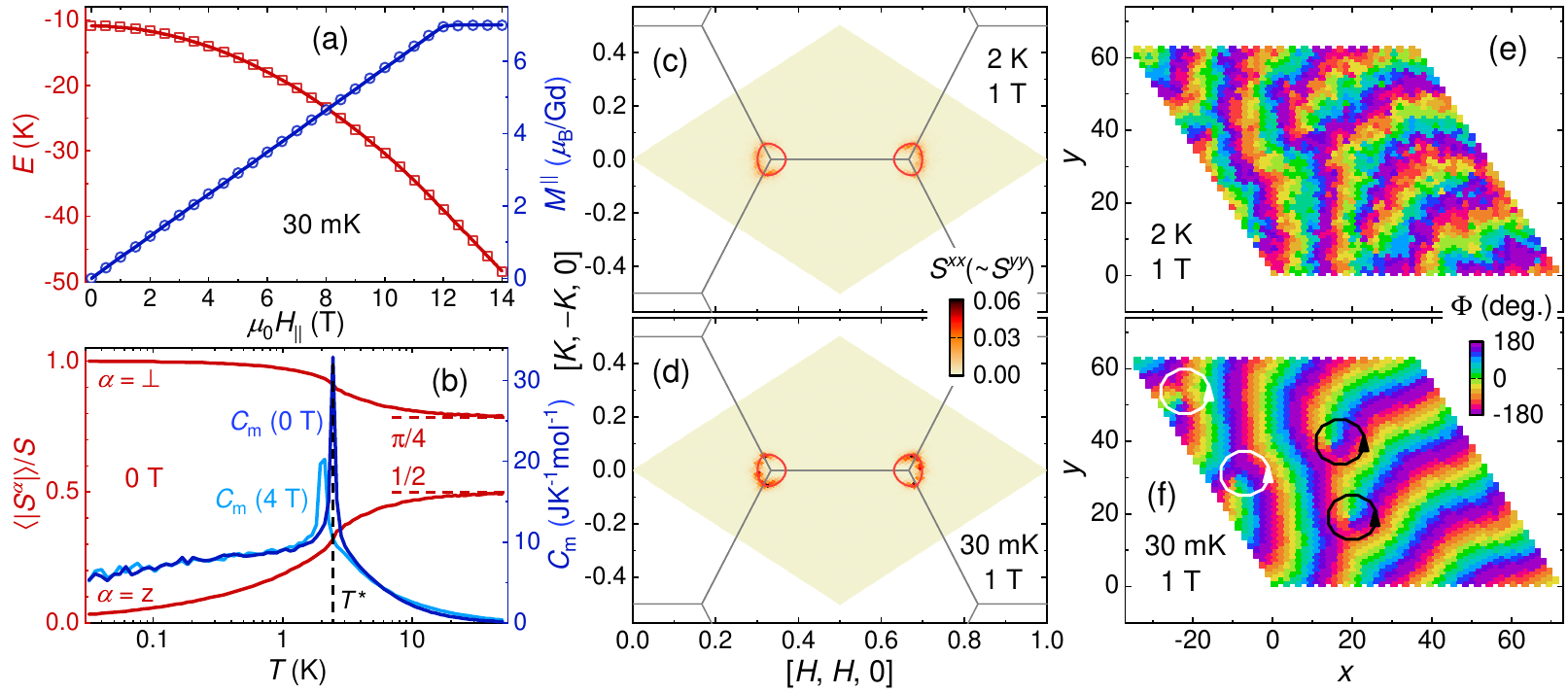}
  \caption{MC simulations for GdZnPO. (a) Energy per spin (red) and magnetization (blue). Scatters represent MC results at 30 mK, while colored lines are analytical results at 0 K using the SSL ansatz. (b) Temperature dependence of average absolute spin components at 0 T (solid red curves: $\langle|S^\perp|\rangle$ $\equiv\langle\sqrt{(S^{x})^2+(S^{y})^2}\rangle$ and $\langle|S^z|\rangle$) and specific heat [dark blue: $C_\mathrm{m}$(0 T); light blue: $C_\mathrm{m}$($\mu_0H^\parallel$ = 4 T)] calculated via MC simulations. Dashed red lines indicate high-temperature limits of $\langle|S^\perp|\rangle$ = $\pi S$/4 and $\langle|S^z|\rangle$ = $S$/2~\cite{supple}. Spin structure factors at 2 K (c) and 30 mK (d) in $\mu_0H^\parallel$ = 1 T. Red lines show the spiral contour; grey lines mark Brillouin zone boundaries. Spin configurations projected onto the $ab$ plane, $\Phi$ = $-i\ln\frac{S^x+iS^y}{\sqrt{(S^x)^2+(S^y)^2}}$, calculated at 2 K (e) and 30 mK (f) in $\mu_0H^\parallel$ = 1 T on sublattice $A$ \uppercase\expandafter{\romannumeral1}~\cite{supple}. White and black circles mark spin vortices [$C_s$ = $\oint\mathrm{d}\Phi(\mathbf{r})/(2\pi)$ = 1] and antivortices ($C_s$ = $-$1), with arrows indicating the integral direction.}
  \label{fig4}
\end{center}
\end{figure*}

\emph{Low-energy spin excitations.}---The magnetic specific heat ($C_\mathrm{m}$) of GdZnPO was accurately determined by subtracting the specific heat of the nonmagnetic YZnPO as the lattice contribution~\cite{supple}. Across all measured fields, the $C_\mathrm{m}$ data show no significant Schottky upturns down to the lowest temperatures [Fig.~\ref{fig3}(a)], and the Gd nuclear contribution is negligible~\footnote{The nuclear specific heat of Gd follows  $C_N$ = $AT^{-2}$, where $A$ slightly varies among compounds. For example, $A$ = 1.37$\times$10$^{-4}$ and 1.75$\times$10$^{-4}$ JK/mol-Gd have been reported for Gd$_2$Sn$_2$O$_7$~\cite{quilliam2007evidence} and Gd$_2$(Ti$_{1-x}$Zr$_x$)$_2$O$_7$ ($x$ = 0.02 and 0.15)~\cite{PhysRevB.83.064403}, respectively. At $T$ $\geq$ 53 mK, $C_N$ $\lesssim$ 0.05 JK$^{-1}$/mol, much smaller than the measured $C_\mathrm{m}$ $\gtrsim$ 1 JK$^{-1}$/mol in GdZnPO, confirming the negligible nuclear contribution.}. At 0 T, $C_\mathrm{m}$ exhibits a peak around $T^*$ $\sim$ 2 K, attributed to the formation of coplanar spin configurations [Fig.~\ref{fig4}(b)] and a crossover between the high-$T$ paramagnetic phase and the SSL, accompanied by the spontaneous breaking of chiral symmetry [Fig.~\ref{fig4}(c)]~\cite{PhysRevResearch.4.013121,supple}. Below $\sim$ $T^*$, $C_\mathrm{m}$ remains large and nearly independent of the longitudinal field at $\mu_0H^\parallel$ $\lesssim$ 8 T. Below $\sim$ 0.5 K, $C_\mathrm{m}$ follows the behavior $C_\mathrm{m}$ $\sim$ $C_0+C_1T$, with $C_0$ = 1.20(4) JK$^{-1}$/mol and $C_1$ = 2.74$\pm$0.16 JK$^{-2}$/mol, for $\mu_0H^\parallel$ $\lesssim$ 8 T. In sharp contrast, at a longitudinal field of 11.8 T (close to $\mu_0H_\mathrm{s}^\parallel$), $C_\mathrm{m}$ drops significantly and approaches zero as $T$ $\to$ 0 K. From the lowest measured temperature $T_\mathrm{min}$ = 53 mK ($\ll$ $|J_1|S^2$) up to $\sim$ 20 K ($\gg$ $|J_1|S^2$), the system releases an entropy [$S_\mathrm{m}(T)$ = $\int_{T_\mathrm{min}}^TC_\mathrm{m}(T')/T'\mathrm{d}T'$] $\sim$ 3$R\ln2$ = $R\ln(2S+1)$ [see the inset of Fig.~\ref{fig3}(a)], indicating the absence of zero-point entropy in GdZnPO. Additionally, the small $S_\mathrm{m}$ suggests that the system approaches its ground states at $T$ $\lesssim$ 0.15 K.

Similarly, the magnetic susceptibility becomes constant at $\chi_0^\parallel$ $\sim$ 5 cm$^3$/mol as $T$ $\to$ 0 K [Fig.~\ref{fig3}(b)]. At low temperatures, the magnetization exhibits full polarization at $H^\parallel$ $\sim$ $H_\mathrm{s}^\parallel$ and nearly linear polarization at $\mu_0H^\parallel$ $\lesssim$ 9 T [inset of Fig.~\ref{fig3}(b)].

\emph{Discussion.}---The measured residual specific heat $C_0$ = 1.20(4) JK$^{-1}$/mol at $\mu_0H^\parallel$ $\lesssim$ 8 T in GdZnPO is unusually large, a phenomenon not previously reported in strongly correlated magnets to our knowledge. For instance, the 3D SSL candidates MnSc$_2$S$_4$ and LiYbO$_2$ exhibit magnetic specific heat $C_\mathrm{m}$ $\sim$ 0.2 and 0.01 JK$^{-1}$/mol, respectively (after removing nuclear contributions), at temperatures as low as $\sim$ 0.1 K~\cite{fritsch2004spin,bordelon2021frustrated}. The 2D SSL candidates Ca$_{10}$Cr$_7$O$_{28}$ and FeCl$_3$ show $C_\mathrm{m}$ $\sim$ 0.16 and 0.02 JK$^{-1}$/mol at $\sim$ 0.037 and 0.6 K, respectively~\cite{PhysRevB.100.174428,tahar2006zero}. Rare-earth spin-ice Dy$_2$Ti$_2$O$_7$, which exhibits significant zero-point entropy, has $C_\mathrm{m}$ $\sim$ 0.044 JK$^{-1}$/mol at $\sim$ 0.4 K~\cite{ramirez1999zero}. The rare-earth gapless spin-liquid candidate YbMgGaO$_4$ shows $C_\mathrm{m}$ $\sim$ 0.2 JK$^{-1}$/mol at 60 mK~\cite{li2015gapless}. These $C_\mathrm{m}$ values are considerably smaller than the $C_0$ observed in GdZnPO. A slightly faster drop in $C_\mathrm{m}$ is visible below $\sim$ 0.15 K at 0 T [Fig.~\ref{fig3}(a)], possibly indicating a precursor to ``order by disorder''~\cite{PhysRevB.81.214419}. However, zero-field $C_\mathrm{m}$ $\sim$ 1 JK$^{-1}$/mol remains large even at 53 mK, and applying a field of $\mu_0H^\parallel$ = 1 or 4 T can prevent the drop down to $\sim$ 65 mK. The rapid decrease in $C_\mathrm{m}$ observed at 8 and 11.8 T below $\sim$ 0.2 K can be attributed to a gap opening due to the Zeeman interaction, as the spin system approaches full polarization.

Due to the presence of the spiral contour, the specific heat behavior of $C_\mathrm{m}$ $\sim$ $C_0+C_1T$ had been predicted for a generic SSL within the spherical approximation~\cite{bergman2007order,yao2021generic}. Moreover, our classical MC simulations also indicate that $C_\mathrm{m}$ $\to$ \emph{constant} as $T$ $\to$ 0 K [see Fig.~\ref{fig4}(b)]. Therefore, the observation of a giant $C_0$ in GdZnPO is distinctive among all existing strongly correlated magnets and should serve as a strong signature of the spiral contour and SSL. Furthermore, the sudden drop of $C_\mathrm{m}$ from $\sim$ 8 to 11.8 T, suggesting that GdZnPO could be utilized in demagnetization cooling to mK temperatures~\cite{tokiwa2021frustrated,xiang2024giant}, especially when a moderate-to-strong magnetic field is unavoidable.

The SSL ansatz is confirmed by unbiased MC calculations for both energy and magnetization, as shown in Fig.~\ref{fig4}(a). The low-$T$ limit of specific heat calculated by the classical model, $C_0^\mathrm{cal}$ $\sim$ 5 JK$^{-1}$/mol [Fig.~\ref{fig4}(b)], is significantly larger than the measured $C_0$ $\sim$ 1.2 JK$^{-1}$/mol. Additionally, the low-$T$ limit of susceptibility calculated using the classical SSL ansatz, $\chi_0^{\parallel,\mathrm{cal}}$ = $\mu_0N_\mathrm{A}g^2\mu_\mathrm{B}^2/[2D+3J_1+9J_2+J_1^2/(4J_2)]$+$\chi_\mathrm{vv}^\parallel$~\footnote{$\chi_\mathrm{vv}^\parallel$ $\sim$ 0.3 cm$^3$/mol at low temperatures.} $\sim$ 4.4 cm$^3$/mol, is smaller than the measured $\chi_0^\parallel$ $\sim$ 5 cm$^3$/mol. These discrepancies at low temperatures may be attributed to the quantization of spin and quantum fluctuations in the real material GdZnPO, despite $S$ = 7/2 being relatively large. Additionally, weak perturbations beyond Eq.~(\ref{eq1}) in the real material could also contribute to these discrepancies.

Next, we examine the stability of the SSL in GdZnPO as the temperature decreases. Neither the specific heat nor the susceptibility shows clear evidence of further magnetic order below $T^*$ (Fig.~\ref{fig3}). As the temperature decreases from $\sim T^*$ to 30 mK, the spin structure factor calculated at the classical level does not exhibit further concentration at special points [compare Figs.~\ref{fig4}(d) with \ref{fig4}(c)], which contrasts sharply with conventional magnetic orders. Our simulations for GdZnPO, consistent with previously reported results~\cite{PhysRevResearch.4.013121}, suggest that ``order by disorder'' is still prevented by thermal fluctuations down to 30 mK ($\sim$ 0.006$|J_1|S^2$), aligning with experimental observations. The occurrence of ``order by disorder'' at lower temperatures remains a possibility and warrants further investigation.

Finally, we examine the low-energy topological defects. The SSL phase in GdZnPO approximates a nearly decoupled three-sublattice 120$^\circ$ coplanar order in the $ab$ plane. Consequently, we decompose the honeycomb lattice into six triangular sublattices: $A$ \uppercase\expandafter{\romannumeral1}, $A$ \uppercase\expandafter{\romannumeral2}, $A$ \uppercase\expandafter{\romannumeral3}, $B$ \uppercase\expandafter{\romannumeral1}, $B$ \uppercase\expandafter{\romannumeral2}, and $B$ \uppercase\expandafter{\romannumeral3}. At low temperatures, spin vortices and antivortices become evident, with spin winding numbers $C_s$ = 1 and $-$1, respectively, as illustrated in Fig.~\ref{fig4}(f). These topological defects persist at $T$ $\sim$ $T^*$, as shown in Fig.~\ref{fig4}(e). The orientations of neighboring spins on different sublattices are locked together, causing the positions of vortices and antivortices to remain fixed across different sublattices. Additionally, the corresponding local (sublattice) momentum vortices~\cite{PhysRevResearch.4.023175} are observed around the positions of spin vortices and antivortices on the sublattices, as shown in Fig.~S6 of~\cite{supple}. Since these sublattices are discernible in the SSL phase below $\sim T^*$, their topological configurations may be utilized in developing small-scale antiferromagnetic spintronic devices~\cite{jungwirth2016antiferromagnetic,RevModPhys.90.015005,gao2020fractional}.

\emph{Conclusions.}---We establish GdZnPO as an ideal candidate for the frustrated honeycomb model, which stabilizes a distinctive SSL with a spiral contour around the $\mathrm{K}$ point. Experimental observations, including the giant residual specific heat $C_0$ $\sim$ 1.2 JK$^{-1}$/mol, provide strong evidence for the formation of SSL down to extremely low temperatures. The distinctive SSL supports the emergence of low-energy spin vortices and antivortices, as well as momentum vortices, on the sublattices at low temperatures. This may be a significant step towards realizing antiferromagnetic spintronic devices.

\acknowledgments
\emph{Acknowledgments.}---We gratefully acknowledge Shang Gao, Zhengxin Liu, Haijun Liao, and Changle Liu for helpful discussion. This work was supported by the National Key R\&D Program of China (Grant No. 2023YFA1406500), the National Natural Science Foundation of China (Nos. 12274153 and U20A2073), and the Fundamental Research Funds for the Central Universities (No. HUST: 2020kfyXJJS054).

\bigbreak

\bibliography{GdZnPO}

\clearpage

\addtolength{\oddsidemargin}{-0.75in}
\addtolength{\evensidemargin}{-0.75in}
\addtolength{\topmargin}{-0.725in}

\newcommand{\addpage}[1] {
 \begin{figure*}
   \includegraphics[width=8.5in,page=#1]{Supplementary_r1.pdf}
 \end{figure*}
}

\addpage{1}
\addpage{2}
\addpage{3}
\addpage{4}
\addpage{5}
\addpage{6}
\addpage{7}
\addpage{8}
\addpage{9}

\end{document}